\documentclass[lettersize,journal]{IEEEtran}
\usepackage{amsmath,amsfonts}
\usepackage{array}
\usepackage{textcomp}
\usepackage{siunitx}
\usepackage{enumitem}
\usepackage{stfloats}
\usepackage{url}
\usepackage{verbatim}
\usepackage{graphicx}
\usepackage{cite}
\usepackage{graphicx}
\usepackage{subfigure} % 导言区

\usepackage{bm}
\usepackage{stmaryrd}
\usepackage{booktabs}
\usepackage{multirow} 
\usepackage{amsthm}

\usepackage{amssymb}
\usepackage{amsmath}
\usepackage{amsmath,amssymb,amsfonts}

\usepackage{siunitx}
\usepackage{xcolor}
\definecolor{deepred}{RGB}{122,1,1}
\usepackage{makecell}
\usepackage{array}

\usepackage{graphicx}
\usepackage{subcaption}

\newtheoremstyle{ieeeremark}
  {3pt}                    
  {3pt}                    
  {\normalfont}             
  {}                       
  {\bfseries}               
  {.}                       
  { }                       
  {}  
  
\theoremstyle{ieeeremark}

\usepackage{amsmath, amssymb, amsthm}

\theoremstyle{plain}
\usepackage[utf8]{inputenc} 
\usepackage[T1]{fontenc}
\usepackage{url}
\usepackage{ifthen}
\usepackage{amsfonts,algorithm,algpseudocode,xcolor}

\begin{document}

%\title{Modulo Sampling: Fixed-Order Difference Recovery with an FPGA-Based Folding Front-End}
%\title{FPGA-Based Modulo Sampling System with ×100 Dynamic-Range Expansion: Hardware Design and Performance Evaluation}
\title{FPGA-Enabled Modulo ADC with ×100 Dynamic-Range Expansion: Hardware Design and Performance Evaluation} 

\author{Zeyuan Li, Wenyi Yan, Lu Gan, Guoquan Li, Hongqing Liu
\thanks{
This work has been submitted to the IEEE for possible publication. Copyright may be transferred without notice, after which this version may no longer be accessible.}
\thanks{}}


\maketitle

\begin{abstract}
Conventional analog-to-digital converters (ADCs) fail to capture high-dynamic-range (HDR) signals due to clipping. Modulo ADCs circumvent this limitation by folding the input prior to quantization and algorithmically reconstructing the original waveform. This work presents a field-programmable gate array (FPGA)-based modulo ADC platform for systematic HDR performance evaluation. The mixed-signal architecture integrates a precision analog front end with a 200-MHz FPGA control loop that incorporates multi-bit updates and digital under-compensation calibration, ensuring stable folding and accurate feedback generation. The system achieves more than a hundred-fold dynamic-range expansion within a 400-kHz bandwidth while maintaining fidelity comparable to that of a conventional ADC. A system-on-chip (SoC)-like implementation enables on-board real-time recovery and supports benchmarking of state-of-the-art reconstruction algorithms, providing a compact and practical framework for HDR signal acquisition and evaluation.
\end{abstract}


\section{Introduction}
Analog-to-digital converters (ADCs) convert analog signals into digital data, serving as the bridge between the physical world and computer systems. They are fundamental components in sensors, communication systems, digital signal processing, and test-and-measurement instruments such as oscilloscopes, spectrum analyzers, and network analyzers~\cite{ADCSensor,ADCApp}. However, conventional ADCs suffer from clipping when high-dynamic-range (HDR) signals exceed the input range, leading to irreversible information loss~\cite{koma_wide_2015, jo_very_2016,guarnieri_high_2011}.

% \begin{figure}[t]
%     \centering
%     \includegraphics[width=0.9\linewidth]{Figures/ModuloADCVsCon.jpg}
% \caption{Comparison of modulo and conventional ADCs for a sinc input. The modulo ADC folds the signal within $[-\lambda,\lambda)$, while the conventional ADC clips at $\pm\lambda$.}
%     \label{fig:modulo_vs_clipping}
% \end{figure}

Modulo sampling systems~\cite{AB_Sampta,bhandari_unlimited_2018,bhandari_unlimited_2021} address this limitation by folding the input signal into a bounded interval $[-\lambda,\lambda)$ prior to digitization, followed by algorithmic reconstruction of the original waveform. 
For a real-valued signal $g(t)$ with a modulo threshold $\lambda>0$, the operation is defined as~\cite{bhandari_unlimited_2021}
\begin{equation}
y(t)=\mathcal{M}_{\lambda}\{g(t)\}
=g(t)-2\lambda\left\lfloor\frac{g(t)+\lambda}{2\lambda}\right\rfloor,
\label{eq:mod}
\end{equation}
where $\lfloor\cdot\rfloor$ denotes the floor function. 
% Fig.~\ref{fig:modulo_vs_clipping} illustrates the preservation of waveform structure by modulo folding, in contrast to the clipping distortion observed in a conventional ADC when applied to a sinc input. 
A key performance metric of modulo ADCs is the dynamic-range expansion factor, defined as
\begin{equation}
\rho \triangleq \frac{\|g\|_\infty}{\lambda},
\end{equation}
where $\|g\|_\infty$ denotes the bounded signal amplitude.
The ratio $\rho$ expresses how many times the input amplitude exceeds the threshold and thus the effective dynamic-range extension.

Numerous algorithms have been proposed for reconstructing modulo-sampled signals. Among them, approaches based on high-order and first-order difference formulations have been theoretically validated as effective examples~\cite{bhandari_unlimited_2021,Guo2023_ICASSP_ITERSIS,ZhuLSE,yan2025differencebasedrecoverymodulosampling}. However, practical hardware implementations capable of verifying their performance under real measurement conditions remain limited, as existing modulo ADC prototypes offer only modest folding depth ($\rho<20$) and bandwidth~\cite{zhu2025ironing,Zhu_TIM_2024}. Existing modulo ADC prototypes are typically limited by their use of analog feedback loops or microcontroller-based control.
Analog implementations suffer from bandwidth constraints, temperature drift, and stability issues, whereas microcontroller-based systems exhibit insufficient response speed for high-frequency folding.
These limitations hinder precise calibration and make systematic performance evaluation difficult.

Seeing the limitations of existing hardware implementations, we develop an field-programmable gate array (FPGA)-based modulo ADC platform as the main control system to provide flexible digital operation, precise calibration, and deeper folding capability.
The main contributions of this work are summarized as follows:
\begin{itemize}
\item \textbf{Deeper and high-frequency folding operation:}
The system adopts a mixed-signal architecture with a 200 MHz FPGA-based digital control loop, supporting up to 400 kHz signal bandwidth and achieving over-hundred-fold dynamic-range expansion ($\rho>100$).

\item \textbf{Digitally calibrated and SoC-like real-time observation:}
The SoC-like mixed-signal architecture incorporates digital calibration to reduce folding errors such as overshoot and integrates an on-board direct reconstruction module, enabling real-time monitoring of folding dynamics and system stability.

\item \textbf{Unified and extensible experimental platform:}
The developed platform supports configurable modulo ADC thresholds and diverse test signals, including both basic and communication waveforms, providing a unified environment for comprehensive evaluations of existing recovery algorithms.
\end{itemize}

% The remainder of this paper is organized as follows. Section~\ref{sec:liter} reviews the background and related work. Section~\ref{sec:hard} presents the FPGA-based folding hardware prototype. Section~\ref{sec:experi} evaluates its performance through simulations and hardware measurements. 

The remainder of this paper is organized as follows.
Section~\ref{sec:liter} summarizes existing hardware implementations and representative recovery algorithms.
Section~\ref{sec:hard} describes the proposed FPGA-based mixed-signal prototype in detail.
Section~\ref{sec:experi} presents hardware measurement results under various signal configurations.
Section~\ref{sec:con} concludes the paper.

\section{Literature Review}
\label{sec:liter}
\subsection{Basics of Modulo Sampling and Conventional Dynamic Range Extension techniques}
% Consider a real bandlimited signal \(g(t)\), which is folded prior to digitization by a modulo operation limiting its amplitude to the interval \([-\lambda,\,\lambda)\). The resulting waveform is expressed as \(y(t) = \mathcal{M}_{\lambda}\{g(t)\}\). It is then uniformly sampled with period \(T = 1/f_s\), yielding the discrete sequence \(y[k] = \mathcal{M}_{\lambda}\!\left(g(kT)\right)\). A digital recovery algorithm reconstructs an estimate of the original samples, denoted by \(\tilde{\gamma}[k]\). Because the ADC processes only bounded inputs, large peaks that would normally cause clipping are instead wrapped by the modulo operation and subsequently unwrapped algorithmically.
Consider a real bandlimited signal \(g(t)\) that is folded prior to 
digitization by a modulo operation restricting its amplitude to 
\([-\lambda,\lambda)\). The folded waveform is denoted by \(y(t)\). 
Ideal noiseless uniform sampling with sampling period \(T = 1/f_s\), 
where \(f_s\) is the sampling rate, yields
\(
y[k] = \mathcal{M}_\lambda\!\left(g(kT)\right).
\)
Digital recovery methods aim to reconstruct the unfolded samples \(g[k]\) 
from the bounded modulo data \(y[k]\), producing an approximation 
\(\tilde{g}[k]\) of the true values.

% In this framework, large excursions of \(g(t)\) that would 
% normally exceed the ADC's dynamic range are wrapped by the modulo operation and 
% later unwrapped algorithmically.

This approach fundamentally differs from classical HDR techniques. \emph{Automatic gain control (AGC)} prevents ADC overload by reducing gain when large signal components appear, but global attenuation can bury weak features below the quantization floor~\cite{bolanos2020_agc_biomed,rovetta2017_agc_ecg,TCAS_agc}. \emph{Oversampling $\Delta\Sigma$ converters} enhance in-band SNR through noise shaping and decimation, proving effective at modest bandwidths, but the required oversampling ratios and decimation complexity scale unfavorably with bandwidth and resolution~\cite{Delta_TCASI, tan2020incremental_tutorial,walden_analog--digital_1999}. \emph{Successive-Approximation-Register (SAR) ADCs} achieves energy efficiency and can extend dynamic range (DR) through range switching and multi-step conversions, yet remain constrained by fixed full-scale limits~\cite{tang2022lowpower_sar_survey,kuo2021multistep_incremental,TCAS_SARADC}. 

In contrast to these approaches that \emph{prevent} signal overload, modulo sampling deliberately accepts localized folding events and shifts DR extension to the \emph{signal processing domain}. 

% Section~\ref{subsec:hardware_impl} reviews existing hardware prototypes, followed by the recovery algorithm in Section~\ref{sec:ms_theory_alg}.

\subsection{Existing Recovery Algorithm}\label{sec:ms_theory_alg}
% By definition of the modulo operation in~\eqref{eq:mod}, the sampled signal can be expressed as~\cite{bhandari_unlimited_2021}
% \begin{equation}
% g[k] = y[k] + \varepsilon_g[k], \quad \varepsilon_g[k] \in 2\lambda\mathbb{Z},
% \end{equation}
% where \(\varepsilon_g[k]\) is the residual function accounting for the integer multiple of \(2\lambda\) introduced by the folding process. 

The modulo samples satisfy the decomposition~\cite{bhandari_unlimited_2021}
\[
g[k] = y[k] + \varepsilon_g[k], \qquad \varepsilon_g[k]\in 2\lambda\mathbb{Z},
\]
where \(\varepsilon_g[k]\) is the discrete version of the residual signal
\(\varepsilon_g(t)\).

% Most single-channel modulo ADC recovery algorithms aim to estimate $\varepsilon_g[k]$ from the folded samples $y[k]$ and add it back to reconstruct $g[k]$. 
Bhandari \textit{et al.}~\cite{bhandari_unlimited_2021} established the Unlimited Sampling framework (USF), 
which employs high-order finite differences to unwrap folded samples. 
When the sampling rate exceeds $1/(2\Omega e)$, the modulo operator becomes inactive, enabling accurate recovery of the residual function. 
Yan \textit{et al.}~\cite{yan2025differencebasedrecoverymodulosampling} further tightened the sampling condition to $T < 1/\Omega$ 
and fixed the difference order to two (RSoD) to enhance reconstruction robustness. 
Subsequent studies adopted first-order formulations for improved stability and computational efficiency: 
Guo \textit{et al.}~\cite{Guo2023_ICASSP_ITERSIS} presented the ITER-SIS algorithm, 
an iterative approach resilient to noise and spectral leakage; 
Shah \textit{et al.}~\cite{Shah2024_LASSO_B2R2} proposed the LASSO-B$^2$R$^2$ method, 
which formulates residual estimation as a sparse recovery problem using LASSO; 
and Zhang \textit{et al.}~\cite{ZhuLSE} introduced the USLSE technique, 
combining dynamic programming and Orthogonal Matching Pursuit (OMP) for line spectral estimation of modulo-sampled signals in the frequency domain.

Beyond the single-channel methods reviewed above, several approaches incorporate additional information to resolve folding ambiguities~\cite{Gan_Liu_MultiADC,GuoIrr,GuoSubNyquist,PVMulti,DorianMulti}.
Bernardo \textit{et al.}~\cite{Neil1bits} proposed recovery schemes with and without auxiliary 1-bit folding information in the presence of quantization noise.
Chinese Remainder Theorem (CRT)-based multi-channel architectures employ multiple modulo ADCs with distinct thresholds to achieve HDR reconstruction~\cite{ICASSP24_MultiChannel,Gan_Liu_MultiADC}.
Guo \textit{et al.}~\cite{GuoSubNyquist} further extended this idea to a quad-channel sub-Nyquist USF system using distinct thresholds and time-delayed channels.

In this work, two recovery modes are considered:   
\textbf{(a)} When the folding count is known and recorded in the FPGA, the original signal is directly reconstructed from \(y[k]\) and the stored count for real-time monitoring and visualization of the folding process, further analyzed in Section~\ref{sec:DRM}.  
\textbf{(b)} When the folding count is unknown, only the folded signal is used, and single-channel recovery algorithms reconstruct the waveform for performance evaluation (Section~\ref{sec:experi}).

\subsection{Prototype Hardware Implementations}\label{subsec:hardware_impl}

Early modulo-sampling hardware prototype employs \emph{purely analog} loops composed of an op-amp integrator and dual comparators for threshold detection and feedback~\cite{Bhandari2022TSP_FP,Florescu2022TSP_Hysteresis,Florescu2022_ICASSP_LocAvg,zhu_60_10Hz,zhu2025ironing}.
The comparators generate polarity-reversed feedback pulses when the integrator output exceeds $\pm\lambda$, folding large signals within a limited voltage range.
This compact, low-power design operates without digital control or clocking but suffers from distortion at high frequencies due to limited analog bandwidth, integrator saturation, and temperature drift.
Guo~\emph{et al.}~\cite{GuoBExpansion} demonstrated that accurate recovery is still achievable by algorithmically compensating folding distortion, albeit with higher computational complexity and dense oversampling requirements.
These limitations motivate mixed-signal architectures integrating analog folding with digital counting and DAC feedback for improved stability and control.

% The prototypes have adopted \emph{mixed-signal architectures} where an analog folding circuit is combined with digital counting and DAC-based feedback to realize real-time modulo sampling~\cite{Zhu_TIM_2024,mulleti_hardware_2023}. 
The mixed-signal architecture typically comprises an analog front end with an operational-amplifier chain, comparators for threshold detection, and a DAC-based feedback path, while a microcontroller handles digital counting and control.  
Zhu~\emph{et~al.}~\cite{Zhu_TIM_2024} demonstrated an analog--digital prototype integrating these components, achieving a dynamic-range expansion of approximately \(\rho \approx 10\) within an \(\sim8~\text{kHz}\) bandwidth.  
Mulleti~\emph{et~al.}~\cite{mulleti_hardware_2023} reported a similar design using a faster Teensy~4.1 controller, reaching about \(\rho \approx 8\) at \(\sim10~\text{kHz}\).
In both systems, the folding depth was limited by the supply voltage, threshold level, and DAC resolution, while the effective bandwidth was constrained by control-loop latency, mainly determined by the digital controller response time, analog front-end bandwidth, and DAC settling time.

This work addresses these limitations with an FPGA-based mixed-signal design enabling deeper and more accurate folding through improved timing and digital calibration.

\begin{figure}
    \centering
    \includegraphics[width=1\linewidth]{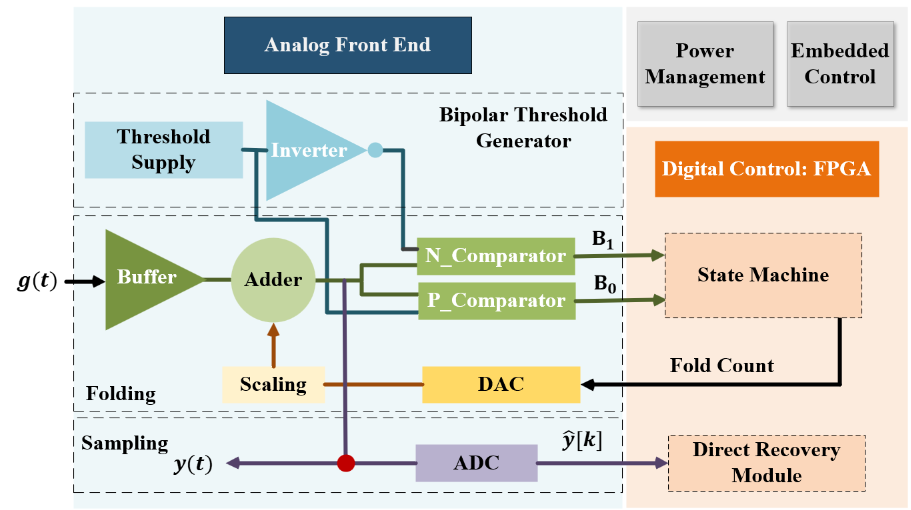}
    \caption{Simplified block diagram of the FPGA-based modulo ADC prototype.}
    \label{fig:system_overview}
\end{figure}

\begin{figure*}[t]
    \centering
    \includegraphics[width=0.7\linewidth]{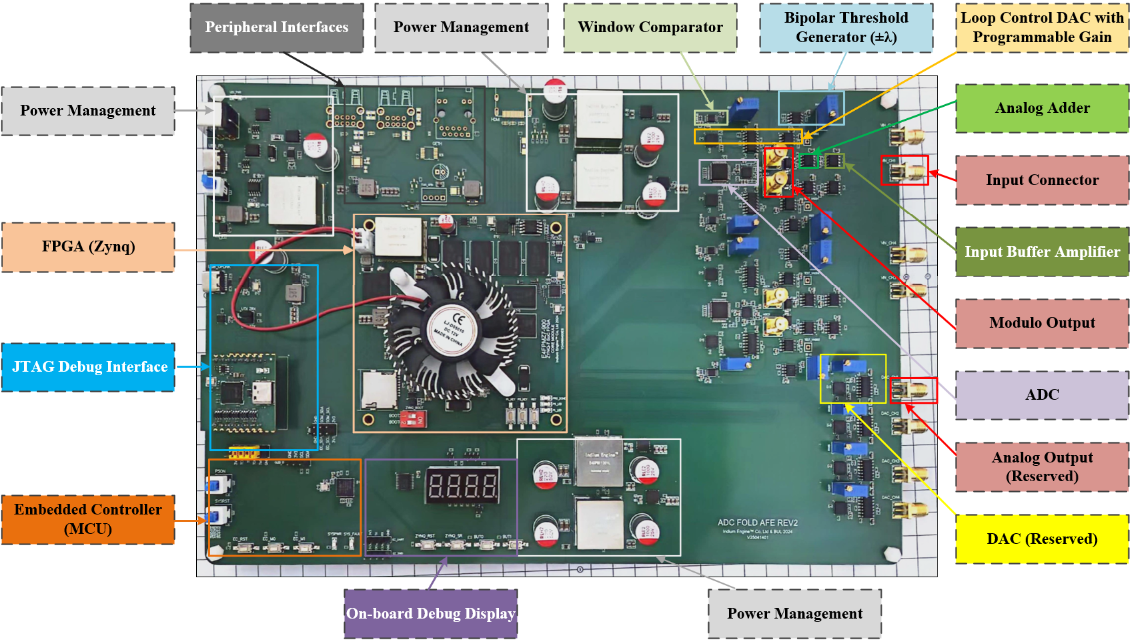}
\caption{Photograph of the fabricated FPGA-based modulo ADC prototype.  
Key components of a single channel are labeled, including the FPGA-based digital loop controller, window comparator, bipolar threshold generator, Loop Control DAC with programmable gain stage, adder, and input buffer.  
Additional modules, such as the on-board ADC/DAC, power management, and peripheral interfaces, are integrated to enable stand-alone operation and future system expansion.}
    \label{fig:hardwarePro}
\end{figure*}

\section{Hardware Implementation} \label{sec:hard}
\subsection{System Overview and Implementation Highlights}

From a signal-flow perspective, Fig.~\ref{fig:system_overview} illustrates the operation of the FPGA-based modulo ADC prototype.  
The input signal \(g(t)\) passes through the analog front end (AFE), where it is continuously monitored by a high-speed window comparator.  
Whenever the input voltage exceeds either threshold \(\pm\lambda\), the comparator asserts digital flags to the FPGA.  
A finite-state machine (FSM) interprets these flags and updates a signed folding-count register \(C_f\) at the FPGA control-loop clock rate \(T_c\).  
This register tracks the cumulative number of modulo folds performed in the analog domain and directly drives the DAC.

To model the feedback path, we denote by \(C_f[n]\) the value of the folding-count register \(C_f\) at the \(n\)-th FPGA clock cycle.  
The DAC performs zero-order-hold (ZOH) reconstruction, holding each value \(C_f[n]\) over the interval \([nT_c,(n+1)T_c)\).  
After passing through the AFE scaling stage with calibrated gain \(G_{\mathrm{total}}\), the resulting analog folding-feedback voltage becomes a continuous-time, piecewise-constant signal that can be written as
\begin{equation}
v_f(t)
=
\sum_{n=-\infty}^{\infty}
\bigl(G_{\mathrm{total}} V_{\mathrm{DAC,step}}\, C_f[n]\bigr)
\,\mathrm{Rect}_{T_c}(t - nT_c),
\label{eq:vf_rect_general}
\end{equation}
where \(V_{\mathrm{DAC,step}}\) is the DAC output increment per fold, and
\(\mathrm{Rect}_{T_c}(t)\) denotes a unit-amplitude rectangular pulse of duration \(T_c\), defined as
\begin{equation}
\mathrm{Rect}_{T_c}(t)
=
\begin{cases}
1, & 0 \le t < T_c, \\
0, & \text{otherwise}.
\end{cases}
\label{eq:rect_def}
\end{equation}

The gain is configured to satisfy 
\(G_{\mathrm{total}} V_{\mathrm{DAC,step}} = 2\lambda\),
so that each update of the folding-count register applies an exact 
\(2\lambda\) correction at the AFE summing node.  Because the DAC 
holds the register value constant over each control-loop interval 
of duration \(T_c\), the feedback voltage reduces to the simple 
piecewise-constant form
\[
v_f(t) = 2\lambda\, C_f \, ,
\quad t \in [nT_c,(n+1)T_c),
\]
where \(C_f\) denotes the current value of the folding-count 
register during that interval.

The folded waveform \(y(t)\) can be observed directly on an oscilloscope or 
digitized by the on-board ADC, producing discrete samples \(\hat{y}[k]\), which
represent the noisy measurements obtained in practice.  
These samples can be viewed in real time through the integrated logic analyzer (ILA) in Vivado.  
A Direct Recovery Module (DRM) within the FPGA reconstructs the signal by 
combining \(\hat{y}[k]\) with the folding count \(C_f\).

% From a signal-flow perspective, Fig.~\ref{fig:system_overview} illustrates the operation of the FPGA-based modulo ADC prototype.  
% The input signal \(g(t)\) passes through the analog front end (AFE), where it is monitored by a high-speed window comparator.  
% When the input voltage exceeds the bipolar thresholds \(\pm\lambda\), the comparator sends digital flags to the FPGA.  
% A finite-state machine (FSM) interprets these flags, updates an internal counter \(C_f\) that records the cumulative folding count, and adjusts the DAC output to generate a \emph{folding feedback} signal \(v_f(t) = -\varepsilon_g(t)=2\lambda\, C_f\), corresponding to the negative residual term.  
% This feedback ensures that the folded analog output \(y(t)\) remains confined within the range \([-\lambda, +\lambda)\). The folded waveform \(y(t)\) can be observed directly on an oscilloscope or digitized by the on-board ADC to produce discrete samples \(y[k]\) stored in the FPGA.  
% These samples can also be viewed in real time through the integrated logic analyzer (ILA) in Vivado.  
% A \emph{Direct Recovery Module (DRM)} within the FPGA reconstructs the input signal by combining \(y[k]\) with the accumulated counter \(C_f\).  
% This on-chip recovery enables signal validation measurement for direct comparison between the modulo ADC output and the reconstructed waveform.

The design features:  
\textbf{(a)} a dynamic-range extension of approximately \(\rho>100\) achieved through precise fold-step control and a digital calibration system;  
\textbf{(b)} wideband, low-latency folding operation with a detection bandwidth exceeding 400~kHz; and  
\textbf{(c)} an integrated, extensible platform with power management and peripheral interfaces supporting both stand-alone and multi-channel operation.  
Fig.~\ref{fig:hardwarePro} shows the four-channel PCB operating from a ±12~V supply, with one active channel highlighted and peripheral circuitry omitted for clarity.  
Table~\ref{tab:hardware-components} summarizes the key hardware specifications.

\renewcommand{\arraystretch}{1.3}
\begin{table}[t]
\centering
\caption{List of hardware components.}
\label{tab:hardware-components}
\begin{tabular}{l|l|l}
\toprule
\hline
\textbf{Component} & \textbf{Model number} & \textbf{Make} \\
\hline
Comparator        & LT1715        & Analog Devices\\
\hline
DAC        & AD9744      & Analog Devices     \\
\hline
PGA& VCA824& Texas Instruments\\
\hline
Adder / Buffer            & OPA892& Texas Instruments\\
\hline 
Inverter            & ADA4522 &  Analog Devices \\
\hline
 ADC& AD9288&Analog Devices\\
\hline
FPGA   & ZYNQ 7000 XC7Z100   & Xilinx               \\
\hline
\bottomrule
\end{tabular}
\end{table}

\subsection{Analog Front-End Design}
The AFE of the proposed modulo ADC is designed to generate, detect, and correct threshold-crossing events with high precision and minimal latency.
In mixed-signal folding architectures, two hardware factors primarily limit system performance:
\textbf{(a)} the supply voltage range, threshold setting, and DAC resolution of the analog circuitry, which together determine the achievable folding depth; and
\textbf{(b)} the total feedback-loop delay, which constrains the folding bandwidth.
Each analog block in the proposed design is carefully selected and optimized to mitigate these limitations.

\subsubsection{Precision Threshold Generation}
\label{subsec:com} 
The proposed design employs a precision bipolar threshold generator and a high-speed window comparator to ensure consistent switching behavior.  
The threshold voltage \(\lambda\) determines both the sensitivity of the folding process and the achievable folding depth, which scales inversely with \(\lambda\) for a fixed supply voltage:  
\[
\text{Folding Depth} \propto \frac{V_{\text{supply}}}{\lambda}.
\]
With \(V_{\text{supply}} = \pm12~\text{V}\), smaller thresholds enable proportionally deeper folding but increase the risk of noise-induced false triggering, whereas larger thresholds improve noise immunity at the cost of reduced folding resolution.  
In this work, a nominal threshold range of \([-0.5,+0.5]\)~V is used, providing a practical trade-off between stability and folding depth.
Specifically, for \(V_{\text{supply}} = 12~\text{V}\) and a threshold of \(\lambda = 0.1~\text{V}\), the theoretical maximum folding depth is \(\rho = 120\) for an analog front-end fully rail-to-rail.

The threshold generator employs a zero-drift amplifier (ADA4522) and a precision voltage reference to generate symmetric \(\pm\lambda\) limits within the desired range.  
These components are chosen for their excellent thermal stability and low offset, maintaining voltage balance within \(\pm0.2\%\) and effectively eliminating temperature-induced drift.  
Threshold detection is performed using a dual-channel high-speed comparator (LT1715) featuring a 4~ns propagation delay, 150~MHz bandwidth, and a 3.5~mV Schmitt-trigger hysteresis.  
The hysteresis suppresses chattering near \(\pm\lambda\), while the matched dual channels ensure synchronized detection of both polarities.  
The comparator outputs are encoded as a 2-bit flag \([B_1B_0]=[\text{OVRN},\text{OVRP}]\), where:
\begin{itemize}
    \item \textbf{01}: Input \(>+\lambda\) (positive over-range),
    \item \textbf{10}: Input \(<-\lambda\) (negative over-range),
    \item \textbf{00}: Input within \([-\lambda,+\lambda)\).
\end{itemize}

\subsubsection{Folding Feedback Generation and Depth Configuration}
Conventional modulo-sampling front-ends rely on analog integrator–comparator
loops~\cite{Bhandari2022TSP_FP,Florescu2022TSP_Hysteresis}.  
While simple, their limited bandwidth caps the folding rate, and drift or mismatch
causes gain and threshold deviations.  
Because the folding step is hardware-fixed, calibration and tuning are difficult.  
This motivates a design with a fully programmable folding threshold.

% To address these limitations, the proposed design implements a high-speed, digitally controlled, and easily calibratable \emph{folding feedback} mechanism.
% In this design, the feedback voltage is given by
% \(
%     v_f(t) = G_{\mathrm{total}} V_{\mathrm{DAC,step}} C_f .
% \)
% The folding threshold \(\lambda\) can be flexibly adjusted, as the folding 
% amplitude depends only on the ratio between \(G_{\mathrm{total}}\) and \(\lambda\). 
% Thus, changing \(\lambda\) merely requires setting the gain to satisfy 
% \(G_{\mathrm{total}} V_{\mathrm{DAC,step}} = 2\lambda\). 
Using~\eqref{eq:vf_rect_general}, the feedback reduces within each clock interval to
\(v_f(t)=G_{\mathrm{total}}V_{\mathrm{DAC,step}}C_f\),
and the folding step is fixed by
\(G_{\mathrm{total}}V_{\mathrm{DAC,step}}=2\lambda\),
so that changing \(\lambda\) only requires adjusting the gain.
In the prototype design, the output path comprises a 14-bit AD9744 DAC 
(200~MSPS, with 1 bit reserved for the sign) whose current output is converted 
to voltage through a Thévenin-equivalent resistor network, followed by a 
programmable-gain amplifier (VCA824, adjustable \(G\in[0,2]~\mathrm{V/V}\)) and 
a fixed \(\times10\) buffer stage.  
This configuration provides a total gain range of 
\(G_{\mathrm{total}}\in[0,20]~\mathrm{V/V}\), a bandwidth of 
\SI{200}{\mega\hertz}, and a slew rate of 500~V/\si{\micro\second}.

% \textbf{Theoretical Folding Depth.}
% In the ideal case, one DAC least significant bit (LSB) after amplification corresponds to a single fold:
% \begin{equation}
%     G_{\mathrm{total}} V_{\mathrm{LSB}} = 2\lambda,
% \end{equation}
% where $V_{\mathrm{LSB}} = V_{\mathrm{FS}} / 2^{13}=V_{\mathrm{DAC,step}}$ represents the voltage step of a 14-bit DAC (one bit reserved for the sign). 
% This configuration yields a maximum theoretical folding depth of
% \begin{equation}
%     M_{\mathrm{fold},\max}^{(\mathrm{ideal})} = 2^{13} = 8192
% \end{equation}
% per polarity, or $16\,384$ folds in total.

\textbf{Practical Limitation and Multi-Bit Step Design.}
To enhance adaptability across different threshold settings $\lambda$ 
while operating within the fixed supply voltage ($\pm12~\mathrm{V}$), 
a programmable multi-bit update scheme is employed, 
where each fold corresponds to $2^q$ DAC codes:
\begin{equation}
    V_{\mathrm{DAC,step}} = 2^q V_{\mathrm{LSB}}.
\end{equation}
where \(V_{\mathrm{LSB}} = V_{\mathrm{FS}} / 2^{13}\) and \(V_{\mathrm{FS}}\) is the full-scale output voltage of the DAC.  The parameter $q$ provides real-time control over the effective DAC resolution, 
ensuring consistent folding symmetry and balanced dynamic-range utilization. Following this design principle, the folding feedback is expressed as  
\begin{equation}
    v_f(t) = G_{\mathrm{total}} 2^q V_{\mathrm{LSB}} C_f.
\end{equation}
%The maximum achievable bipolar folding depth is given by  
The achievable range of bipolar folding depth is
\begin{equation}
C_{f} \in[- 2^{13-q},  2^{13-q}-1]
\end{equation}
For example, to achieve a dynamic-range expansion of \(\rho = 108\), the system 
must satisfy \(\rho/2 < C_{f,\max}\).  
Choosing \(q = 7\) gives \(C_{f,\max} = 63\), which meets this requirement.  
Although smaller values of \(q\) also satisfy the inequality, the largest 
admissible \(q\) is selected to ensure efficient utilization of the DAC 
resolution without wasting bits.

\subsubsection{Wideband Summation}
The real-time summation of the input and feedback signals is implemented using a unity-gain summing amplifier.  
With a 500~MHz gain--bandwidth product, 500~V/$\mu$s slew rate, and $< 40$~ns settling time, the OPA892 provides high linearity and wide bandwidth.

% \textbf{Comparison with existing work.}
% Zhu~\emph{et~al.}~\cite{Zhu_TIM_2024} used a 3-bit resistor-DAC driven by a microcontroller, where coarse quantization caused feedback saturation at $\pm15$~V, limiting the folding depth to $K_{\max}=7$.
% Mulleti~\emph{et~al.}~\cite{mulleti_hardware_2023} employed a similar low-resolution DAC, achieving $K_{\max}\approx8$.
% In contrast, the proposed design uses a 14-bit high-linearity DAC enabling deeper folding.

\subsection{Digital Loop Controller}
The FPGA serves as the central digital controller that manages folding detection, counter updates, and folding voltage compensation under a unified clock domain.

\subsubsection{Folding Controller}
The FPGA-implemented FSM (\SI{200}{\mega\hertz}) processes AFE comparator outputs to control the fold count \(C_f\). It receives the 2-bit status \([B_1B_0]\) from Section~\ref{subsec:com} and wait flag \(B_2\) (indicating settling completion).

\begin{figure}[t]
    \centering
    \includegraphics[width=0.9\linewidth]{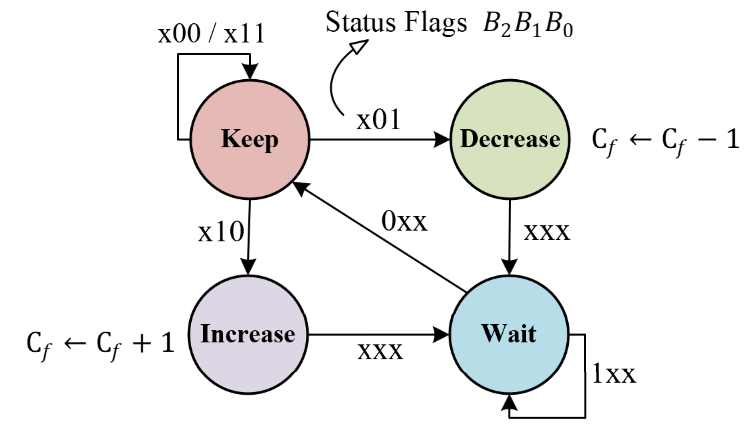}
\caption{Control loop FSM: States \textbf{KEEP}/\textbf{WAIT} (hold $C_f$), \textbf{INCREASE}/\textbf{DECREASE} ($C_f \!\pm\! 1$). Transitions depend on $[B_2 B_1 B_0]$ and ``x'' means don't-care.\vspace{-0.3cm}}
    \label{fig:StateMachine}
\end{figure}

The FSM (Fig.~\ref{fig:StateMachine}) operates through four states:
\begin{itemize}
    \item \textbf{KEEP}: Hold \(C_f\), \([B_2B_1B_0]=[\text{x}00]/[\text{x}11]\), where \(x\) denotes a don't-care bit.
    \item \textbf{INCREASE}: \(C_f \leftarrow C_f+1\) if \([B_2B_1B_0]=[\text{x}10]\).
    \item \textbf{DECREASE}:\(C_f \leftarrow C_f-1\) if \([B_2B_1B_0]=[\text{x}01]\).
    \item \textbf{WAIT}: Keep \(C_f\) unchanged while \([B_2=1]\), allowing AFE settling and preventing oscillation.
\end{itemize}

\subsubsection{System Clocking and Synchronization}
To ensure precise synchronization among the ADC, Folding controller, and loop-control DAC, a unified clock network is implemented using an FPGA-integrated phase-locked loop (PLL).
A 50~MHz reference is multiplied inside the FPGA to generate phase-aligned clocks for all subsystems.
The ADC is driven by a 100~MHz clock advanced by 60$^{\circ}$ to align sampling with the settled feedback signal, while the controller and DAC operate on 200~MHz in-phase clocks.
This configuration provides deterministic timing and stable high-speed operation for the modulo ADC system.

\subsubsection{Direct Recovery Module}  \label{sec:DRM}
FPGA integrates a \emph{direct recovery module} for on-chip reconstruction of the original input signal.  
It combines the stored folding count \(C_f\) with the sampled modulo signal \(\hat{y}[k]\), acquired from the on-board ADC after the analog front end and thus subject to practical errors such as quantization, jitter, and offset.  
The recovered waveform is computed as  
\begin{equation}
    \tilde{g}[k] = \hat{y}[k] - 2\lambda\,C_f,
\end{equation}
enabling real-time reconstruction and direct hardware verification of folding accuracy and dynamic-range expansion, which will be further analyzed in Section~\ref{sec:experi}.

% \begin{figure}[t]
%     \centering
%     \includegraphics[width=0.9\linewidth]{Figures/ClockDistributionNetwork.png}
% \caption{Clock Distribution Network Functional Diagram}
%     \label{fig:ClockDistributionNetwork}
% \end{figure}

\textbf{Comparison with existing work.}
Previous prototypes using STM32 or Teensy controllers relied on interrupt-driven routines, limited by software latency and sequential task execution~\cite{Zhu_TIM_2024,mulleti_hardware_2023}.
Microsecond-level response delays led to missed folding events and inaccurate counter updates at high input frequencies.
The lack of a shared clock among the microcontroller unit (MCU), ADC, and DAC further caused phase misalignment and folding jitter.
In contrast, the FPGA implements all control logic in parallel hardware pipelines, enabling clock-synchronous response and deterministic timing.

\subsection{Folding Calibration}
During measurements, transient overshoot was observed at the folding edges, appearing as short voltage spikes beyond the threshold range.
These spikes arise from abrupt voltage discontinuities in the folding process, which excite reflections along the feedback path.
Because of the finite bandwidth of the amplifier and DAC, the reflected components are partially amplified, producing overshoot at the analog front end.
This overshoot distorts the modulo signal and consequently degrades the accuracy of subsequent signal reconstruction.

To suppress this effect, a \textit{controlled under-compensation} strategy was introduced. 
In the ideal case, after $C_f$ folding events, the folding feedback $v_f(t)$ should follow~\eqref{eq:feed}. 
In practice, however, the actual feedback voltage is intentionally designed to be slightly smaller to suppress overshoot caused by abrupt voltage transitions:
\begin{equation}
    v_{\mathrm{raw}}(t) = C_f\,G_{\mathrm{total}}\,2^{q}V_{\mathrm{LSB}} < v_f(t).
\end{equation}
The feedback deviation is defined as
\begin{equation}
    v_f(t) - v_{\mathrm{raw}}(t)
    = C_f\,\Delta V,
\end{equation}
where
\(
    \Delta V = 2\lambda - G_{\mathrm{total}}\,2^{q}V_{\mathrm{LSB}}.
\)
This difference grows linearly with the folding count. 
To compensate for the accumulated error, the FPGA adds a digital correction term $(C_f-1)\Delta V$, resulting in a calibrated residual
\begin{equation}
    v_{\mathrm{cal}}(t) = v_{\mathrm{raw}}(t) + (C_f-1)\Delta V
    = v_f(t) - \Delta V,
\end{equation}
which remains constant regardless of the number of folds.

\begin{figure}
    \centering
    \includegraphics[width=0.9\linewidth]{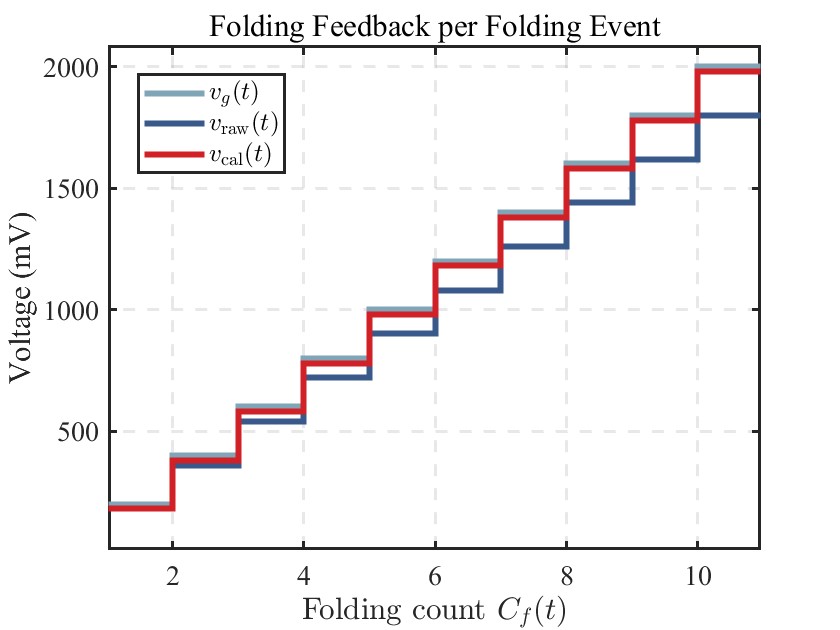}
\caption{
Folding feedback voltage for $\lambda=100~\mathrm{mV}$ with $C_f$ varying from 1 to 10:
$v_f(t)=C_f200~\mathrm{mV}$ (ideal),
$v_{\mathrm{raw}}(t)=C_f180~\mathrm{mV}$ ($\Delta V=20~\mathrm{mV}$, uncalibrated),
and $v_{\mathrm{cal}}(t)=v_f(t)-20~\mathrm{mV}$ (calibrated).
}
    \label{fig:residual_comparison}
    \vspace{-0.5cm}
\end{figure}

Fig.~\ref{fig:residual_comparison} illustrates the digital calibration process 
of the controlled under-compensation scheme. In this simulation, the threshold 
is set to \(\lambda = 100~\mathrm{mV}\), yielding the ideal folding feedback 
\(v_f(t) = C_f\,2\lambda = C_f\,200~\mathrm{mV}\). To suppress transient 
overshoot, the hardware feedback step is intentionally reduced to 
\(v_{\mathrm{raw}}(t) = C_f\,180~\mathrm{mV}\), corresponding to an empirically 
chosen under-compensation margin of \(\Delta V = 20~\mathrm{mV}\). Without 
calibration, this discrepancy accumulates linearly with \(C_f\). After digital 
correction, the calibrated feedback \(v_{\mathrm{cal}}(t) = v_f(t) - 20~\mathrm{mV}\) 
closely matches the ideal response, ensuring stable folding across varying \(C_f\).

\subsection{Bandwidth Limitation}
\begin{figure}[t]
    \centering
    \includegraphics[width=1\linewidth]{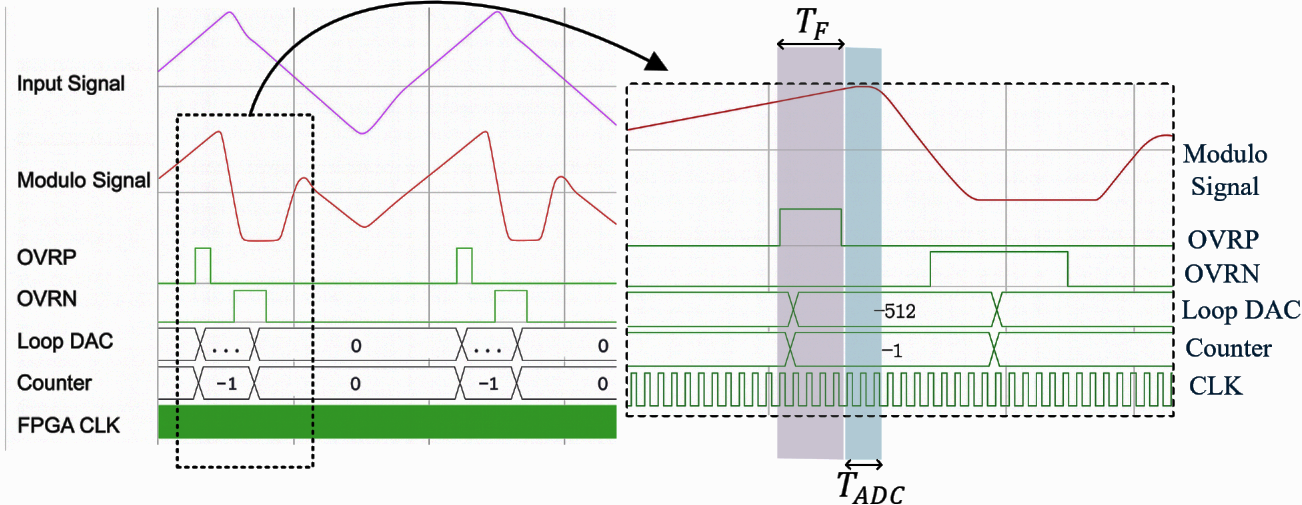}
\caption{Timing-limited folding for a \SI{610}{\kilo\hertz} triangular input with threshold $\lambda=\SI{0.36}{\volt}$. 
The full trace is shown (left) and a zoom near the threshold crossing is shown (right). 
Traces include the input (purple), modulo output (red), OVRP/OVRN flags, DAC step, fold counter, and the \SI{200}{\mega\hertz} clock. 
Loop delay results in an overfold below $-\lambda$, causing spurious OVRN and counter errors.}
  \label{fig:loop_stability}
\end{figure}

The folding bandwidth is limited by both analog and digital delays.  
On the analog side, the finite slew rates of the DAC and summing amplifier slow
the folding transition, whereas the comparator’s \SI{150}{\mega\hertz}
bandwidth is sufficient.  
Digitally, the FPGA response time determines how quickly a correction can be
applied; at high input slew rates, the loop reacts too slowly to keep the signal
within the threshold range.

To characterize these limits, a \SI{610}{\kilo\hertz} triangular input with
$\lambda=\SI{0.36}{\volt}$ was applied, and internal traces were captured
(Fig.~\ref{fig:loop_stability}).  
The signals include the input, modulo output, OVRP/OVRN flags, the DAC update
step ($-2^{9} = -512$ codes), the fold counter $C_f$, and the
\SI{200}{\mega\hertz} clock.  
The loop delay $T_F$, defined as the interval between the rising and falling
edges of OVRP, is five FPGA cycles (\SI{25}{\nano\second}).  
There is also an additional latency $T_{\text{ADC}}$ between the OVRP falling
edge and the actual modulo transition.
At \SI{610}{\kilo\hertz}, the combined delay $T_F + T_{\text{ADC}}$ allows the
input to continue rising after OVRP assertion, introducing a mismatch between
the ideal and measured folding edges.  
Because the transition is no longer sharp relative to the input period, the
delayed correction ($C_f \!\leftarrow\! C_f - 1$) arrives after the slope
reversal, driving the modulo output below $-\lambda$.  
This \emph{over-fold} corrupts the fold count and induces oscillation.

\section{Real Experiments} \label{sec:experi}
This section presents a comprehensive evaluation of the proposed modulo ADC system, including comparisons with a conventional ADC, verification of its SoC-like signal acquisition capability, and experimental validation of existing recovery algorithms using real folded hardware data. 
\begin{figure}[t]
    \centering
    \includegraphics[width=0.9\linewidth]{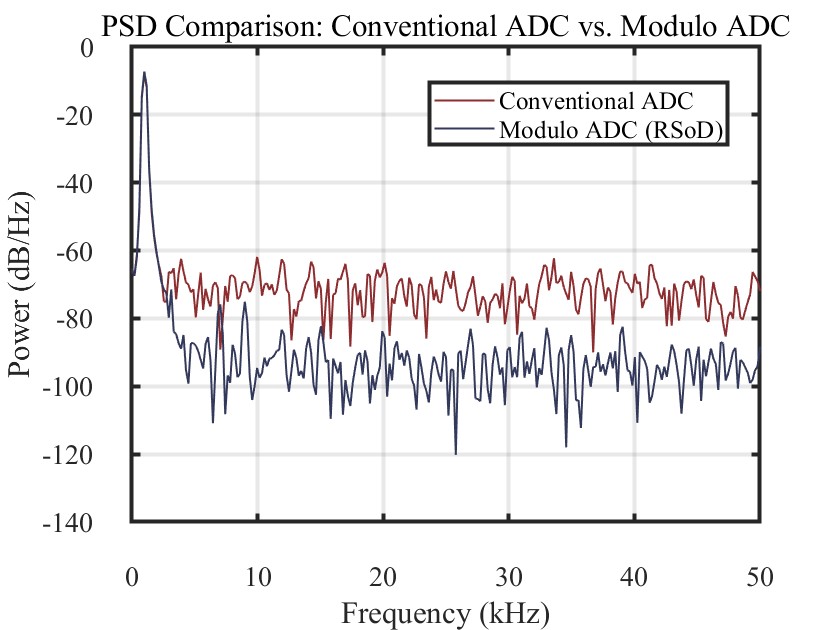}
\caption{Power spectral density (PSD) of the recovered \SI{1}{\kilo\hertz} sinusoid with a resolution of $b = 7~\text{bits/sample}$.} 

% The conventional ADC exhibits a higher noise floor, whereas the modulo ADC with RSoD reconstruction lowers it by approximately \SI{20}{\decibel} over the \SIrange{5}{45}{\kilo\hertz} band.}
    \label{fig:psd_sine}
\end{figure}

\begin{figure}[t]
    \centering
    \includegraphics[width=0.9\linewidth]{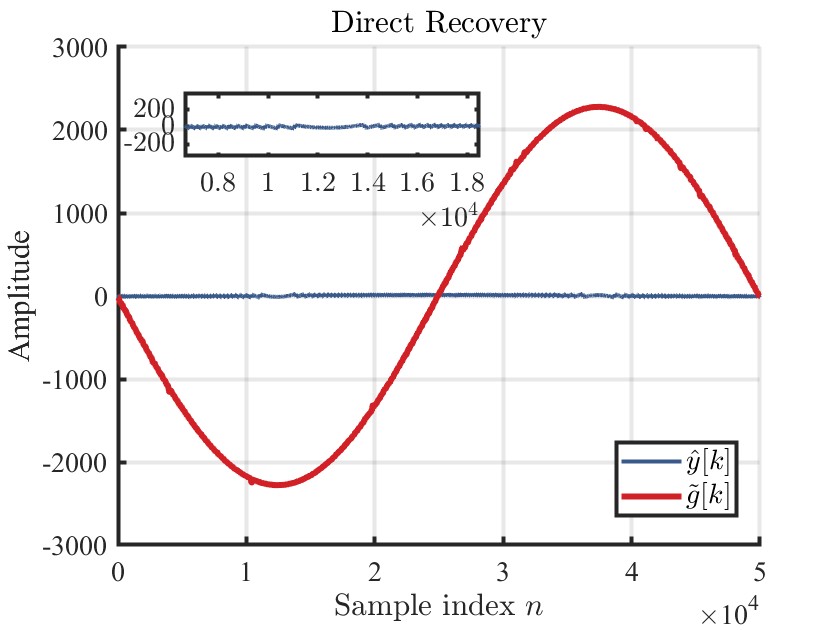}
    \caption{Measured modulo signal \(\hat{y}[k]\) and directly recovered waveform \(\tilde{g}[k]\) at \(\rho = 91.56\) with a modulo threshold of \(0.1~\mathrm{V}\). }
    \label{fig:direcExample}
\end{figure}

\subsection{Comparison with a Conventional ADC} 
Following~\cite{Bhandari2022TSP_FP,Guo2023_ICASSP_ITERSIS,zhu_60_10Hz}, the folded outputs were recorded using a Tektronix TDS1012C-EDU 8-bit oscilloscope.  
The on-board ADC was bypassed to allow software-controlled oversampling and to avoid ADC-induced gain errors during MATLAB-based reconstruction and analysis.

In this experiment, the conventional ADC operated at full scale, with its threshold set equal to the maximum input amplitude to ensure no clipping under nominal conditions.  
For comparison, the modulo ADC was configured with a threshold of \(\lambda=\SI{0.1}{\volt}\) and a dynamic-range factor of \(\rho=108\), using a \SI{1}{\kilo\hertz}\ input.  
The RSoD recovery method~\cite{yan2025differencebasedrecoverymodulosampling} was adopted for reconstruction.  
Fig.~\ref{fig:psd_sine} shows the corresponding power spectral density (PSD) with ADC resolution \(b=7\).  
The conventional ADC exhibits a higher noise floor, whereas the modulo ADC with RSoD reconstruction reduces it by approximately \SI{20}{\decibel} over the \SIrange{5}{45}{\kilo\hertz} band, demonstrating improved noise suppression and reconstruction fidelity.

\subsection{On-Board Direct Signal Recovery}
This experiment aims to verify the direct on-board recovery capability of the proposed modulo ADC system.  The folded waveform was acquired by the on-board AD9288 ADC and stored in the FPGA memory in real time.
As shown in Fig.~\ref{fig:direcExample}, the modulo threshold was set to \(\pm0.1~\mathrm{V}\), corresponding to \(\pm25\) in digital representation.

At $\rho = 91.56$, the reconstructed signal achieves a signal-to-noise and distortion ratio (SINAD) of 44~dB, which is nearly identical to the native 46~dB SINAD of the AD9288 operating within its linear range. This demonstrates that the proposed modulo ADC architecture preserves signal integrity while effectively extending the dynamic range far beyond that of the original ADC. The close SINAD consistency confirms the precise timing alignment, accurate feedback calibration, and high quality of the folded signal, validating the robustness of the overall hardware design.

\begin{table*}[t]
\centering
\caption{Performance comparison for basic test signals (sine and periodic sinc) under various $\rho$, bandwidths $B$, and repetition frequencies $f_m$, with $\lambda = 0.1~\mathrm{V}$. Metrics: OF (dark red = smallest) and SNR$_r$ (bold = highest).}
\renewcommand{\arraystretch}{1.3}
\begin{tabular}{c|c|c|cc|cc|cc|cc}
\toprule
\hline
\multicolumn{1}{c|}{\begin{tabular}[c]{@{}c@{}}Input Signal  Type\end{tabular}}& \multicolumn{2}{c|}{\begin{tabular}[c]{@{}c@{}}Input Signal Parameters\end{tabular}}
& \multicolumn{2}{c|}{RSoD~\cite{yan2025differencebasedrecoverymodulosampling}} 
& \multicolumn{2}{c|}{ITER-SIS~\cite{Guo2023_ICASSP_ITERSIS}} 
& \multicolumn{2}{c|}{LASSO-B$^2$R$^2$~\cite{Shah2024_LASSO_B2R2}} & \multicolumn{2}{c}{USLSE~\cite{ZhuLSE}} \\
\hline 
 & $\rho$ & $f_m$ (\si{\kilo\hertz})  
& $\mathrm{OF}$ & SNR$_{r}$(\si\decibel)& $\mathrm{OF}$  & SNR$_{r}$(\si\decibel)& $\mathrm{OF}$  & SNR$_{r}$(\si\decibel)& $\mathrm{OF}$  & SNR$_{r}$(\si\decibel)\\ \cline{2-11}
\multirow{3}{*}{Sine} & 2.84 & 100  & \textcolor{deepred}{6.49} & 21.62 & 11.11 & \textbf{29.44} & 8.33 & 23.38 & 7.04  & 22.44 \\  \cline{2-11}
& 22.2 & 10  & \textcolor{deepred}{17.86} & \textbf{22.57} & 71.42 & 20.24 & 71.42 & 22.30 & 62.5  & 20.54 \\  \cline{2-11}
&102 & 1   & \textcolor{deepred}{41.67} & \textbf{23.37} & 62.5 & 6.37 & 50 & 0.49 & 50  & 0.33 \\ \hline
\multirow{4}{*}{Sinc} & $\rho$ & $B, f_m$ (\si{\kilo\hertz})  
& $\mathrm{OF}$ & SNR$_{r}$(\si\decibel)& $\mathrm{OF}$  & SNR$_{r}$(\si\decibel)& $\mathrm{OF}$  & SNR$_{r}$(\si\decibel)& $\mathrm{OF}$  & SNR$_{r}$(\si\decibel)\\  \cline{2-11}
& 3.24 & 410, 23 (\si{\kilo\hertz})   & 3.81 & \textbf{24.31} & \textcolor{deepred}{2.97} & 23.86 & 3.93 & 22.22 & 3.48 & 21.76 \\  \cline{2-11}
& 9.16 & 99, 5 (\si{\kilo\hertz}) & \textcolor{deepred}{7.01} & 23.09 & 11.48 & \textbf{27.79} & 12.63 & 22.78 & 12.63 & 24.31 \\  \cline{2-11}
& 29.80 & 18, 1 (\si{\kilo\hertz}) & \textcolor{deepred}{15.43} & 16.43 & 46.30 & \textbf{17.13} & 46.30 & 16.07 & 46.30 & 13.47\\
\hline
\bottomrule
\end{tabular}
\label{tab:alg_comparison_sinc}
\end{table*}

\begin{table*}[t]
\centering
\caption{Performance comparison of RSoD~\cite{yan2025differencebasedrecoverymodulosampling}, ITER-SIS~\cite{Guo2023_ICASSP_ITERSIS}, LASSO-B$^2$R$^2$~\cite{Shah2024_LASSO_B2R2}, and USLSE~\cite{ZhuLSE} using communication-type test signals with different $B$ and $\rho$. Metrics: OF (dark red = smallest) and SNR$_r$ (bold = highest).}
\renewcommand{\arraystretch}{1.3}
\begin{tabular}{c|c|c|cc|cc|cc|cc}
\toprule
\hline
\multicolumn{3}{c|}{\begin{tabular}[c]{@{}c@{}}Input Signal Parameters\end{tabular}}
& \multicolumn{2}{c|}{RSoD~\cite{yan2025differencebasedrecoverymodulosampling}} 
& \multicolumn{2}{c|}{ITER-SIS~\cite{Guo2023_ICASSP_ITERSIS}} 
& \multicolumn{2}{c|}{LASSO-B$^2$R$^2$~\cite{Shah2024_LASSO_B2R2}} & \multicolumn{2}{c}{USLSE~\cite{ZhuLSE}} \\
\hline 
 Type & $\rho$ & $B$ (\si{\kilo\hertz})  
& $\mathrm{OF}$ & SNR$_{r}$(\si\decibel)& $\mathrm{OF}$  & SNR$_{r}$(\si\decibel)& $\mathrm{OF}$  & SNR$_{r}$(\si\decibel)& $\mathrm{OF}$  & SNR$_{r}(\si\decibel)$\\ \hline
QAM & 10.40 & 4  & \textcolor{deepred}{12.50} & 28.06 & 25.00 & \textbf{34.70} & 25.00 & 23.10 & 25.00 & 20.88 \\ \hline
BPSK & 5.20 & 2  & \textcolor{deepred}{17.86} & 23.44 & 22.73 & \textbf{33.64} & 22.73 & 23.39 & 19.23 & 23.55 \\ \hline
FSK & 8.00 & 2  & \textcolor{deepred}{20.83} & 23.24 & \textcolor{deepred}{20.83} & \textbf{33.76} & \textcolor{deepred}{20.83} & 22.63 & 41.67 & 21.48\\ \hline
\hline
\bottomrule
\end{tabular}
\label{tab:alg_comparison_c}
\end{table*}

\begin{figure*}[htbp]
    \centering
    \subfigure[]{
        \includegraphics[width=0.3\linewidth]{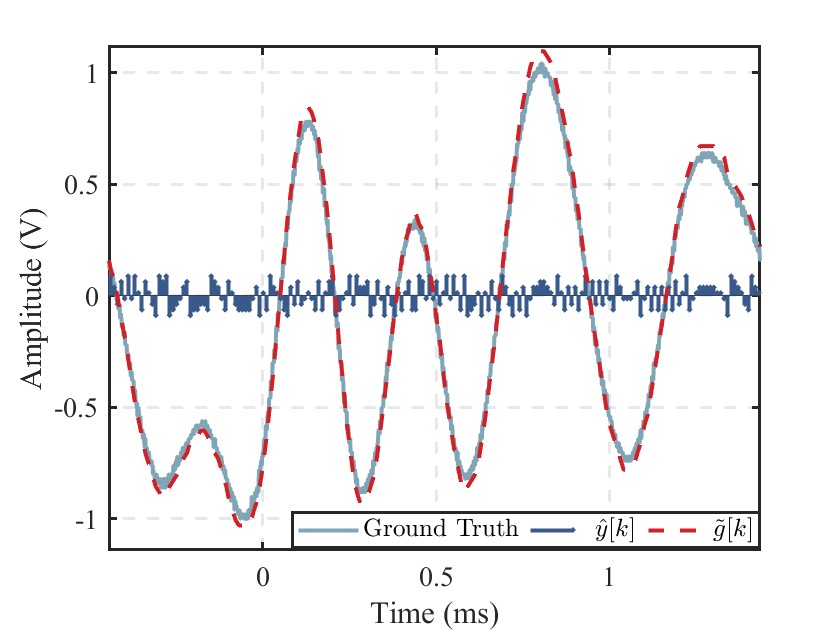}
    }
    \subfigure[]{
        \includegraphics[width=0.3\linewidth]{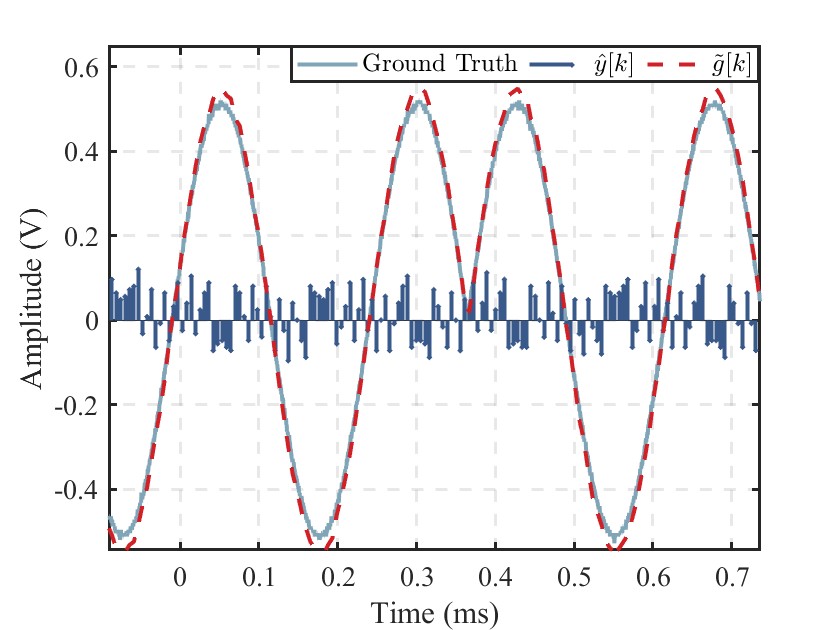}
    }
    \subfigure[]{
        \includegraphics[width=0.3\linewidth]{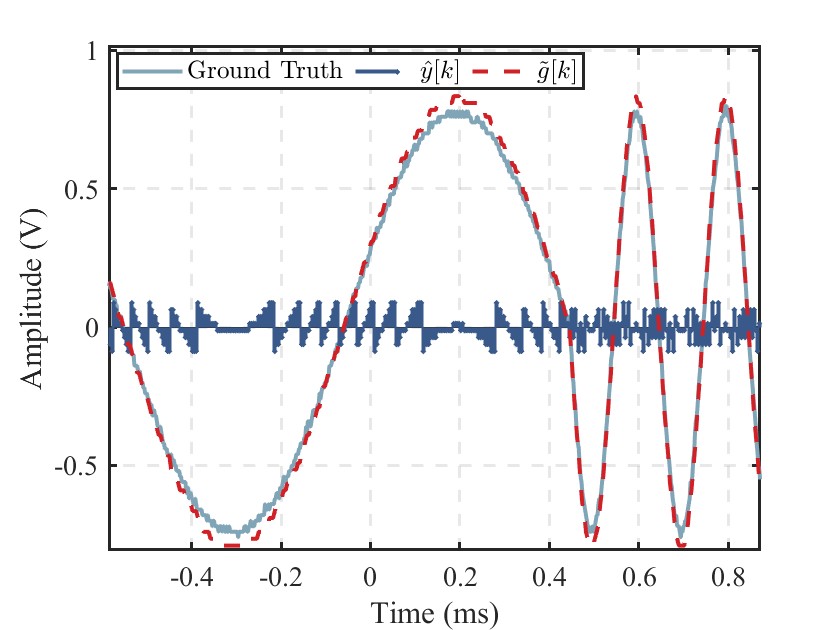}
    }
\caption{Reconstruction results of communication signals using RSoD~\cite{yan2025differencebasedrecoverymodulosampling}: (a) QAM, (b) BPSK, and (c) FSK.}
    \label{fig:c_view}
\end{figure*}

\subsection{Comparison with Existing Recovery Algorithm}
We benchmark against RSoD~\cite{yan2025differencebasedrecoverymodulosampling}, ITER-SIS~\cite{Guo2023_ICASSP_ITERSIS}, LASSO–B$^2$R$^2$~\cite{Shah2024_LASSO_B2R2} and USLSE~\cite{ZhuLSE} under the reconstructed signal-to-noise ratio (SNR$_r$):
\begin{equation}
\text{SNR}_r = 10 \log_{10} 
\frac{\sum_{k} |g[k]|^2}{\sum_{k} g[k]-\tilde{g}[k]|^2}.
\label{eq:rrse_def}
\end{equation}
The ground truth and the folded output were recorded using a Tektronix TDS,1012C-EDU 8-bit oscilloscope.

\subsubsection{Basic Test Signals}
We tested the existing algorithms using basic signals, including single-tone sine waves and periodic bandlimited sinc inputs, each configured with different dynamic ranges $\rho$, signal frequencies $f_m$, and main-lobe bandwidths $B$. The performance metrics are the oversampling factor (OF), defined as $\text{OF} = f_s / f_\mathrm{Nyq}$, and SNR$_r$ defined in~\eqref{eq:rrse_def}. Here, $f_\mathrm{Nyq}$ denotes the Nyquist frequency. The modulo ADC threshold $\lambda$ is fixed at $\lambda = 0.1~\mathrm{V}$.

As summarized in Table~\ref{tab:alg_comparison_sinc}, at low dynamic ranges ($\rho < 10$), all methods yield SNR$_r$ above 20~dB. As $\rho$ increases, ITER-SIS and LASSO–B$^2$R$^2$ require significantly larger sampling factors to sustain reconstruction quality, whereas RSoD preserves similar accuracy with up to threefold reduction in OF. USLSE exhibits stable but moderate performance across all cases. At high dynamic ranges ($\rho > 100$), RSoD maintains SNR$_r$ around 23~dB, while iterative and optimization-based approaches degrade rapidly.

\subsubsection{Communication-Signal Experiments}
To further evaluate the reconstruction performance of the proposed system under practical signal conditions, we tested several typical communication-modulated waveforms, including quadrature amplitude modulation (QAM), binary phase-shift keying (BPSK), and frequency-shift keying (FSK) signals.
As summarized in Table~\ref{tab:alg_comparison_c}, RSoD exhibits the smallest oversampling factors across all QAM and BPSK test cases while maintaining comparable reconstruction quality to the more computationally intensive ITER-SIS and LASSO-B$^2$R$^2$ algorithms. 
Fig.~\ref{fig:c_view} presents the reconstruction results of the communication-signal experiments using the RSoD.

\section{Conclusion and Future Work} \label{sec:con}

This paper presents a mixed-signal FPGA-based modulo ADC platform that achieves more than a 100-fold dynamic-range expansion within a 400~kHz bandwidth and incorporates an under-compensation calibration scheme.  
The on-board direct reconstruction demonstrates that the proposed design achieves a SINAD of 44~dB, comparable to a conventional ADC, while extending the measurable amplitude range by two orders of magnitude.  
Benchmark experiments using both basic and communication signals further show that all evaluated algorithms perform similarly at moderate~$\rho$, whereas only RSoD maintains reliable reconstruction at large~$\rho$.

Future work will focus on improving loop bandwidth and latency control to support higher-frequency inputs.  
The multiple on-board modulo ADC channels will also be exploited to enable synchronized multi-channel acquisition and to further evaluate multi-channel recovery algorithms.

%\newpage
\bibliographystyle{IEEEtran}
\bibliography{refs}

\vfill

\end{document}